\documentclass[11pt,english,aps,prc,showpacs,letter,longbibliography]{revtex4-1}

\usepackage{amsmath}
\usepackage[cmyk]{xcolor}
\usepackage[    bookmarks,
                 bookmarksopen = true,
                 bookmarksnumbered = true,
                 linktocpage,
                 colorlinks = true,
                 linkcolor = purple,
                 urlcolor  = blue,
                 citecolor = magenta,
                 anchorcolor = green,
                 hyperindex = true,
                 hyperfigures]
        {hyperref}

\usepackage{enumitem}
\usepackage{graphicx}
\usepackage{esint}
\usepackage[utf8]{inputenc}
\usepackage[T1]{fontenc}
\usepackage{float}
\usepackage{fancyhdr}
\usepackage{ulem}
\usepackage{lineno}
\usepackage{listings}

\DeclareFixedFont{\ttb}{T1}{txtt}{bx}{n}{8} 
\DeclareFixedFont{\ttm}{T1}{txtt}{m}{n}{8}  

\usepackage{color}
\definecolor{deepblue}{rgb}{0,0,0.5}
\definecolor{deepred}{rgb}{0.6,0,0}
\definecolor{deepgreen}{rgb}{0,0.5,0}

\newcommand\pythonstyle{\lstset{
language=Python,
basicstyle=\ttm,
otherkeywords={self},             
keywordstyle=\ttb\color{deepblue},
emph={MyClass,__init__},          
emphstyle=\ttb\color{deepred},    
stringstyle=\color{deepgreen},
frame=tb,                         
showstringspaces=false            %
}}

\lstnewenvironment{python}[1][]
{
\pythonstyle
\lstset{#1}
}
{}

\newcommand{\etc}{{\textit{etc.}}}
\newcommand{\ie}{{\textit{i.e.}}}

\newcommand{\ba}{\begin{eqnarray}}
\newcommand{\ea}{\end{eqnarray}}

\makeatletter
\usepackage{soul}

\makeatother

\begin{document}

\title{Solving Einstein equations using deep learning}

\author{Zhi-Han Li$^{1}$, Chen-Qi Li$^{1}$, Long-Gang Pang$^{1}$\footnote{email: lgpang@ccnu.edu.cn} 
}

\address{$^{1}$Key Laboratory of Quark \& Lepton Physics (MOE) and Institute of Particle Physics, Central China Normal University, Wuhan 430079, China}


\begin{abstract}

Einstein field equations are notoriously challenging to solve due to their complex mathematical form, with few analytical solutions available in the absence of highly symmetric systems or ideal matter distribution. However, accurate solutions are crucial, particularly in systems with strong gravitational field such as black holes or neutron stars. In this work, we use neural networks and auto differentiation to solve the Einstein field equations numerically inspired by the idea of physics-informed neural networks (PINNs). By utilizing these techniques, we successfully obtain the Schwarzschild metric and the charged Schwarzschild metric given the energy-momentum tensor of matter. This innovative method could open up a different way for solving space-time coupled Einstein field equations and become an integral part of numerical relativity.

\end{abstract}

\keywords{Deep learning, Auto differentiation, Einstein Equation}

\pacs{}

\maketitle

\section{Introduction}
Einstein field equations are a set of coupled partial differential equations which are highly non-linear. Their space-time coupling and complex mathematical form require extensive and intricate calculations for obtaining solutions. Due to the difficulty of finding analytical solutions, the numerical relativistic approach strives to solve these equations using numerical methods to determine the time evolution behaviour of corresponding physical systems, but suffers from several numerous difficulties, including equation formulation issues, boundary condition treatment and computational stability, \etc~In the early 1960s, Hahn and Lindquist\cite{HAHN1964304} carried out the first computer simulation of binary black holes. They ran the simulation for a few dozen time steps, but due to significant errors, the program terminated prematurely. In the subsequent decades, many researchers explored various aspects of the problem\cite{1975PhDT1,1975PhDT2,tb37076x,PhysRevD.14.2443,PhysRevLett.41.1085,PhysRevLett.55.891}. Nonetheless, satisfactory results have remained elusive, primarily attributed to the instability of numerical relativity and the limitations of computational power. However, in the early 1990s, spurred on by the LIGO project, more researchers and funding were devoted to the numerical relativity. Over the next decade, significant progress was made in several areas\cite{19901,19902,19903,19904,19905,19906,19907,19908,19909,199010,199011,199012,199013,199014,199015,199016,199017,199018,199019,199020,199021,199022,199023,199024,199025,199026,199027,199028,199029,199030,199031,199032,199033,199034,199035}, including initial conditions, equation formulations, \etc~A breakthrough came in early 2005, when Pretorious\cite{PhysRevLett.95.121101} achieved a complete simulation of  binary black hole merger. In the same year, independent research groups at the University of Texas at Brownsville (UTB)\cite{PhysRevLett.96.111102} and NASA's Goddard Space Flight Center\cite{PhysRevLett.96.111101} discovered a new method called "moving punctures" that successfully simulated black hole mergers.  With the improved stability of numerical relativity, researchers then shifted their focus to computational efficiency and accuracy. Since then the field of numerical relativity has flourished\cite{Centrella:2010mx}, particularly in the study of the  binary compact object mergers. Notably, the construction of gravitational wave templates based on numerical relativity has played an essential role in the experimental detection of gravitational waves\cite{science361}, as the result of the merger of binary compact systems.

With the wide application of deep learning in astrophysics, we find that many researchers use deep learning to solve astrophysical problems\cite{Kessler:2016uwi,DES:2015bqp,Sanders:2014uva,Wong:2020ise,Lanusse:2017vha,PerreaultLevasseur:2017ltk,Fluri:2018hoy}, especially in the gravitational wave research\cite{George:2017pmj,Zevin:2016qwy,George:2016hay,Wong:2020yig,Green:2020hst,George:2018awu,Vajente:2019ycy,Chua:2019wwt,Razzano:2018fxb,Wei:2019zlc,Dax:2021tsq,Shen:2019vep}.  
The field of numerical relativity is undoubtedly complex, making it a daunting task for beginners to enter. Currently, the majority of numerical solutions to Einstein's field equation involve a 3+1d space-time decomposition, finding the evolution of $\gamma_{ij}$(induced metric) or $K_{ij}$(extrinsic curvature) instead of $g_{\mu\nu}$ or $\partial_k g_{\mu\nu}$,  and subsequent solution using finite difference or spectral method. However, we are excited to explore a different approach in solving the Einstein field equations through the use of deep neural networks and auto differentiation. This method, known as physics-informed neural networks (PINNs), has been widely used in physics to solve complex partial differential equations (PDEs) \cite{1711.06464,2,PINN,Karniadakis2021-nq,2107.09443} and many-body Schrödinger equations \cite{JPSJ.87.074002,KEEBLE2020135743,1909.08423,sci, Pu2023jae,Adams:2020aax,Shi:2021qri,Yang:2022rlw}. PINNs utilizes a deep neural network to represent the solutions of PDEs. Since the network parameters are random numbers before training, the solution may lead to large residuals in PDEs, as well as in initial and boundary conditions. These residuals are used as the loss function of the deep neural network, which can be reduced gradually during training through optimization. In comparison to traditional methods, PINNs do not require discretizing space-time into grids or designing specific formulas to approximate differential operators. PINNs is mesh-free and the utilized auto differentiation provides analytical precision, making it ideal for multi-physics, multi-scale\cite{LEUNG2022111539, 2109.09444,2002.08235,2304.01670,2304.01670}, and space-time coupled problems, even amazing solution speed (such as s Navier-Stokes equations in fluid dynamics\cite{2010.08895}) and extremely complex equations (such as MHD \cite{Rosofsky:2023dtc,2305.07940,Rosofsky:2022lgb,1912.11073,2211.14266,2102.01447} and  turbulence problems \cite{Karpov:2022tro,2211.08579,1912.11073}).

To investigate the potential of physics-informed neural networks (PINNs) in solving the Einstein field equations, we concentrate on extracting the metric tensor $g_{\mu\nu}$ based on a given distribution of matter. In the Einstein field equation, the left-hand side is a function of the metric tensor, while the right-hand side is the energy-momentum tensor (\ie~in the Eq.~\ref{Einstein_field_equation}). For a given distribution of matter, the equations reduce to functions of the metric tensor alone. We represent the metric tensor $g_{\mu\nu}$ by a deep neural network whose inputs are the $x_{\mu}$. The residuals of the Einstein field equations are then incorporated into the loss function as physical constraints for training. Through this approach, we have successfully generated the Schwarzschild metric and the charged Schwarzschild metric, which provide compelling evidence of the effectiveness of deep learning in solving the Einstein field equations.

\section{Method}
In the Schwarzschild spacetime, we adopt the natural unit and assume a spherically symmetric gravitational source with mass $M$. When we seek to the solution of the metric field outside the spherically symmetric gravitational source, the Birkhoff theorem tells us that the vacuum spherically symmetric metric must be static. The metric can therefore be expressed as follows,
\begin{align}
    ds^2 & = -f(r)dt^2 + g(r)dr^2 + r^2(d\theta^2 +sin^2\theta d\varphi^2) \notag \\
    & = g_{00}dt^2 + g_{11}dr^2 + r^2(d\theta^2 +sin^2\theta d\varphi^2)
    \label{metric_expression}
\end{align}
The Einstein field equation is as follows,
\begin{equation}
    R_{\mu\nu} - \frac{1}{2}g_{\mu\nu}R = \kappa T_{\mu\nu}
    \label{Einstein_field_equation}
\end{equation}
where the Ricci tensor $R_{\mu\nu}$ and the Ricci scalar R represent the curvature of spacetime, $\kappa= {8 \pi G \over c^4}$, the energy-momentum tensor $T_{\mu\nu}$ contains the matter sources, and $\mu, \nu$ = 0,1,2,3. By convention, the index = 0  selects a “ time ” component, and index = 1, 2, 3 selects a “space” component. $R_{\mu\nu}$ and R depend on the first and second derivatives of the metric tensor $g_{\mu\nu}$. 

In the vacuum, the energy-momentum tensor of a material field is zero, so Eq.~\ref{Einstein_field_equation} can be reduced to,
\begin{equation}
\label{Eq.3}
    R_{\mu\nu} = 0
\end{equation}
The Christoffle symbols and the Ricci tensors are calculated from the metric components given by Eq.~\ref{metric_expression},
\begin{equation}
    \Gamma_{\mu\nu}^{\alpha}={1 \over 2}g^{\alpha\lambda}(\partial_{\nu}g_{\mu\lambda}+\partial_{\mu}g_{\nu\lambda}-\partial_{\lambda} g_{\mu\nu})
    \label{christoffle_symbols_expression}
\end{equation}
\begin{equation}
    R_{\mu\nu}=\partial_{\lambda} \Gamma_{\mu\nu}^{\lambda}-\partial_{\nu}\Gamma_{\mu\lambda}^{\lambda}+\Gamma_{\sigma\lambda}^{\lambda}\Gamma_{\mu\nu}^{\sigma}-\Gamma_{\nu\sigma}^{\lambda}\Gamma_{\mu\lambda}^{\sigma}
    \label{Ricci_tensor_expression}
\end{equation}

By substituting the Ricci tensor into Eq.~\ref{Eq.3} in the weak field linear approximation, one obtains the analytical solution of the Schwarzschild metric field,
\begin{equation}
    ds^2 = -(1-\frac{2M}{r})dt^2 + (1-\frac{2M}{r})^{-1}dr^2 + r^2(d\theta^2 +sin^2\theta d\varphi^2) 
    \label{Schwarzschild_metric_solution}
\end{equation}
When the gravitational source carries a charge Q, the energy-momentum tensor becomes,
\begin{align} 
    T^{\mu }_{\nu}= \begin{bmatrix} {-\frac{Q^2}{8 \pi r^4}}& 0 & 0 & 0 \\ 
    0 & {-\frac{Q^2}{8 \pi r^4}} & 0 & 0 \\ 
    0 & 0 & {\frac{Q^2}{8 \pi r^4}} & 0 \\ 
    0 & 0 & 0 & {\frac{Q^2}{8 \pi r^4}} \end{bmatrix}
    \label{energy-momentum _tensor_of_charged}
\end{align}  
The analytical charged Schwarzschild solution (Reissner-Nordstrom solution) is obtained in the same way, according to Eq.~\ref{Einstein_field_equation},
\begin{equation}
    ds^2 = -(1-\frac{2M}{r} + \frac{Q^2}{r^2})dt^2 + (1-\frac{2M}{r} + \frac{Q^2}{r^2})^{-1}dr^2 + r^2(d\theta^2 +sin^2\theta d\varphi^2) 
    \label{charged_Schwarzschild_metric_solution}
\end{equation}

\begin{figure}[H]
    \centering 
    \includegraphics[width=1\textwidth]{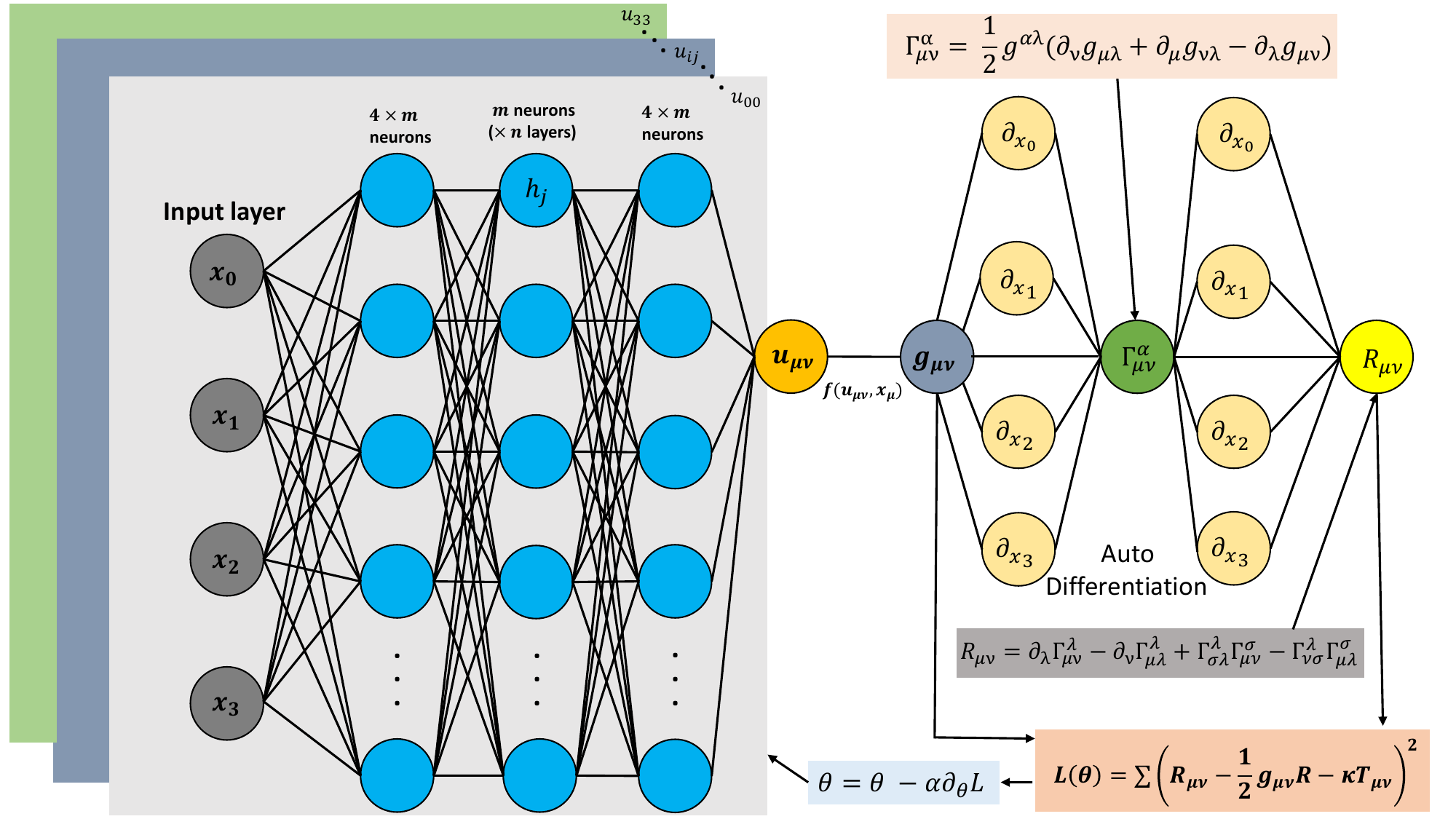}
    \caption{A schematic diagram shows how to solve the metric by deep neural networks. Left panel is the metric $g_{\mu\nu}$ represented by a deep neural networks with space-time coordinates as inputs. Through automatic differentiation showed in the right panel, we can get the Ricci Tensor as an important part in the loss function. The network for metric is trained by optimizing the loss function $L(\theta)$.}
    \label{NN}
\end{figure}

To solve the metric field numerically, we can represent each component of the metric tensor $g_{\mu\nu}$ by one deep neural network, as shown in Fig.~\ref{NN}. To encode some physical information, we don't use the output $u_{\mu\nu}$ of the network as the metric component directly, but construct some functions $f(u_{\mu\nu}, x_{\mu})$, which usually include some boundary conditions to represent the metric.

In our Schwarzschild examples, two deep neural networks are used with the outputs $u_0(r)$ and $u_1(r)$, and the two metric components are constructed as follow,
\begin{equation}
g_{00} = \frac{u_0(r)}{r^2} +\frac{2M}{r} - 1, \quad
g_{11} = u_1(r)
\end{equation}
The current function form of $g_{00}$ encodes the physical constraint from the boundary condition $g_{00} \rightarrow -(1-\frac{2M}{r})$ at $r \rightarrow \infty$, obtained by a linear approximation of the weak gravitational field.

The objective of the training is to find the metric minimizing the destruction to the Einstein field equations (Eq.~\ref{Einstein_field_equation}). The loss function is thus set to be,
\begin{equation}
    L(\theta) = \frac{1}{N} \sum_{i = 0 }^N {(R_{\mu\nu}- \frac{1}{2}g_{\mu\nu} R- \kappa T_{\mu\nu})}^2
\end{equation}
where $\theta$ represent all the trainable parameters in the deep neural networks $u_0(r)$ and $u_1(r)$, $N$ is the number of coordinates uniformly distributed in a given radius interval. The residual $(R_{\mu\nu}- \frac{1}{2}g_{\mu\nu} R- \kappa T_{\mu\nu})$ is computed at each coordinate $r_i$. In areas where there is no material distribution, the loss function is reduced to a simpler form,
\begin{equation}
    L(\theta) = \frac{1}{N} \sum_{i = 0 }^N R_{\mu\nu}^2
\end{equation}

 Mini-batch stochastic gradient descent (mini-batch SGD) is employed to train this deep neural network. The training data are divided into mini batches with equal size $m$. The average loss of one mini batch at a time is used to update the network parameters $\theta$ in gradient descent, by setting $N=m$ in the loss $L(\theta)$,
\begin{equation}
\theta = \theta - \alpha {\partial L \over \partial \theta}
\end{equation}
where $\alpha$ is the learning rate, the gradients ${\partial L \over \partial \theta}$ is computed through auto differentiation \cite{1502.05767}. In practice, we use Adam optimizer \cite{1412.6980} which additionally considers the momentum mechanism and the adaptive learning rate to skip some local saddle points and accelerate the training process. The relevant parameters in the Adam algorithm are set to $\beta_1 = 0.9,\beta_2 = 0.99,\epsilon = 10^{-8}, lr = 10^{-3} $. 
In practice, we need to adjust the learning rate dynamically. A large learning rate in the early stage makes the function converge faster, while a small learning rate makes the training process smooth in the late stage. To make the loss function decrease more steadily in the later stage, we can also increase the batch size at late time. Since the weight and bias parameters of the deep neural network are randomly initialised, the initial output of the neural network may violate the boundary condition and cause the training to diverge (in case $u_0(r)$ contains $r^n,\ n > 2$ components). To prevent the divergence, we follow the same procedure as introduced in the PyTorch deep learning framework \cite{1912.01703}, by initializing the weight parameters using uniformly distributed random numbers in the range $(-\frac{1}{\sqrt{n_o}}, \frac{1}{\sqrt{n_o}})$, where $n_o$ is the number of output neurons in each layer. This distribution has a high probability of producing a first output of the deep neural network that satisfies the boundary condition.

The auto differentiation is also employed to compute the derivatives terms in the Ricci tensor, such as $\partial_{\nu}g_{\mu\lambda}$ and $\partial_{\alpha} \Gamma_{\mu\nu}^{\lambda}$. Automatic differential programming has gained significant popularity as a research area in recent years, offering a distinct approach that differs from both traditional numerical and symbolic differentiation. Numerical differentiation introduces uncontrollable numerical error, while symbolic differentiation often results in complex and opaque expressions. In contrast, automatic differentiation combines the advantages of fast numerical differentiation and the accurate results of symbolic differentiation, enabling the generation of numerical derivatives by accumulating values during code execution. Automatic differentiation has gained popularity in the era of artificial intelligence, thanks to deep learning libraries, such as TensorFlow and PyTorch, which incorporate automatic differentiation functionality. Automatic differentiation computes the derivative $f'(x)$ from a user-defined function $f(x)$ by adding a dual component $x \rightarrow x + \dot{x} \mathbf{d}$ (where $\mathbf{d}$ is a symbol standing for a infinitesimal number satisfying $\mathbf{d}^2 = 0$). By implementing the differentiation of some basic arithmetic operations on a computer, along with chain rules, the user-defined function will automatically have an auto-differentiation dual term, e.g., $f(x) \rightarrow f(x) + f'(x)\mathbf{d}$. PyTorch\cite{1912.01703} is utilized to generate each metric component and calculate their derivatives with respect to the space-time variable using the autograd function. Afterwards, each component of the Christoffel symbols is obtained according to Eq.~\ref{christoffle_symbols_expression}. The derivatives of the Christoffel symbols with respect to the space-time variables are given by auto differentiation, and finally be used to compute each component of the Ricci tensor according to Eq.~\ref{Ricci_tensor_expression}.

Representing the complex metric field over the whole domain with a single neural network makes the learning process very slow. 
The loss decrease too slowly to achieve the desired accuracy at late stage.
The metric field saturate with a significant error (about $10^{-1} \sim 10^{-2}$) near the gravitational source, even with more sampling points added locally.
To overcome this difficulty, we divide the computational domain into smaller pieces, and use different neural networks to represent the metric field in each sub-domain.
This approach simplifies the representation of the metric field by the neural networks, leading to smoother training and more accurate result. 
This distributed computational approach has another name Distributed Physics Informed Neural Network (DPINN) \cite{1907.08967,article111,JAGTAP2020113028}. 
We divide the interval (10, 300) into (10, 30), (30, 80) and (80, 300) for the Schwarzschild metric, and (10, 30), (30, 80), (80, 150) and (150, 300) for the charged Schwarzschild metric field.
The division is an empirical operation, which may lead to discontinuities at the boundaries of sub-domains. 
In practice, the results are good and robust against different choice of dividing schemes. 

In simple feed forward neural network, the inputs are transformed into outputs of each layer nonlinearly through $\sigma(x W + b)$ where $\sigma$ is the activation function. 
Another problem we encountered during the training process was the choice of the activation function. We observe that in the present study, most activation functions (such as tanh, sigmoid, silu) led to poor training accuracy, only LogSigmoid activation brings the training process extremely smooth and efficient.

\section{Results}

\begin{figure}[H]
    \centering 
    \includegraphics[width=1\textwidth]{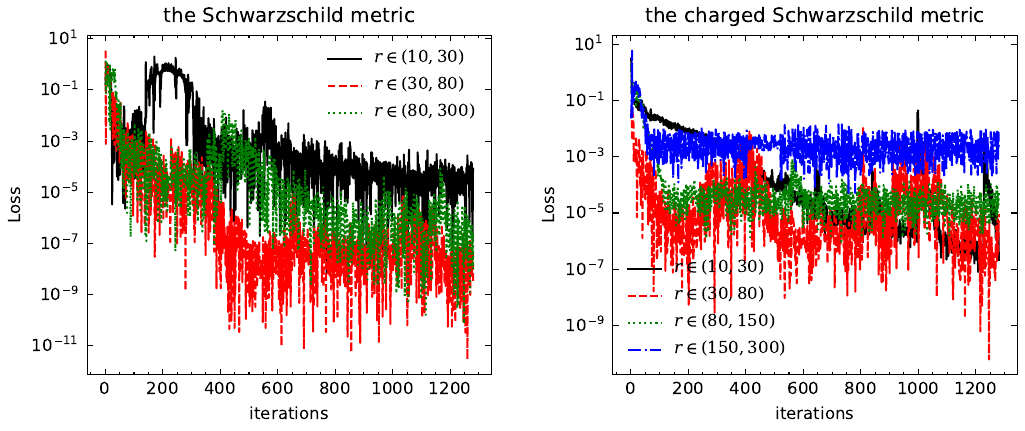}
    \caption{Left: the loss function in the process of training for the Schwarzschild metric. Right: the loss function in the process of training for the charged Schwarzschild metric.}
    \label{fig:loss}
\end{figure}

Fig.~\ref{fig:loss} shows the loss as a function of training iterations, for the Schwarzschild metric (left) and the charged Schwarzschild metric (right). 
The loss of all the neural networks used to represent the metric fields in different regions decrease with time and saturate quickly.
We observe that to achieve desired precision, we have to use more hidden neurons per layer and more training points per volume to train the neural networks that are closer to the gravitational source. In each neural network, we have used five hidden layers.
For the Schwarzschild metric field, we used two deep neural networks with 256 hidden neurons per hidden layer to express the metric components in the region $r\in (10, 30)$, while 128 neurons per hidden layer in $r\in (30, 80)$, and 64 neurons per hidden layer in $r\in (80, 300)$. For the charged Schwarzschild metric field, we use one additional neural network with 32 neurons per hidden layer for the range $r\in (150, 300)$. The other hyperparameters are set to batches = 128, batch size = 8, epoch = 10, which implies that we used 1024 points on every sub-domain. The program takes less than one hour to run on a modern laptop.

\begin{figure}
    \centering
    \includegraphics[width=1\textwidth]{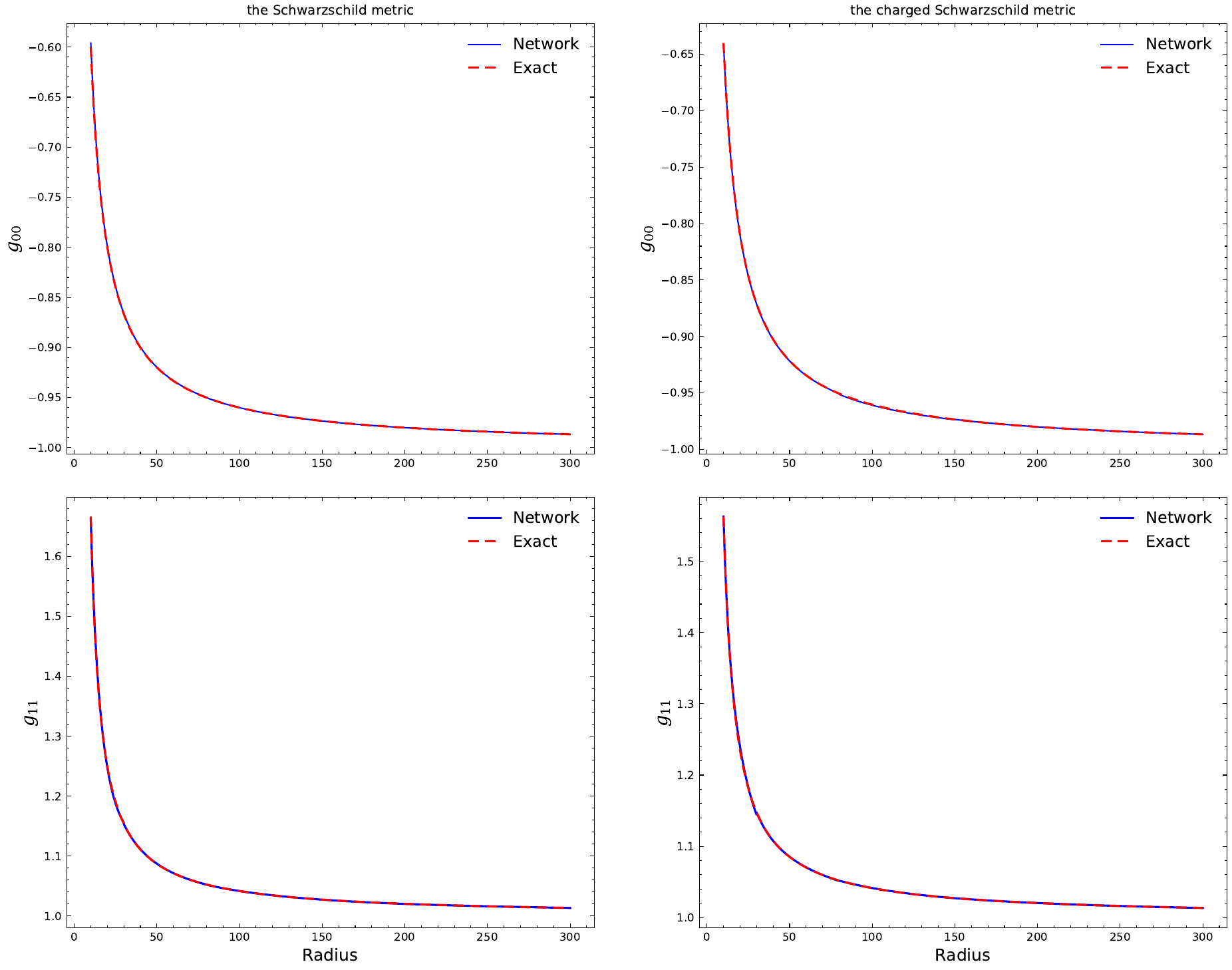}
    \caption{Left: the network solution as compared to the exact result of the Schwarzschild metric in Eq.~\ref{Schwarzschild_metric_solution}. Right: the network solution as compared to the exact result  of the charged Schwarzschild metric in Eq.~\ref{charged_Schwarzschild_metric_solution}}
    \label{fig:metric}
\end{figure}

Shown in Fig.~\ref{fig:metric} is the numerical solutions of Einstein field equations using deep neural networks as compared with analytical solutions, for the Schwarzschild metric field and the charged Schwarzschild metric respectively. 
For the Schwarzschild metric, we set $M = 2$. 
For the charged Schwarzschild metric, we have used $M = 2$ and $Q = 2$, where the energy-momentum tensor (\ie~Eq.~\ref{energy-momentum _tensor_of_charged}) is a non-linear function of the radius.
As described before, the region $r\in(10, 300)$ is divided into 3 and 4 sub-domains for these 2 different cases, 
a different neural network is used to represent the metric for each domain,
while discontinuities between different domains are negligible.    
The metric fields learned by the neural network agree well with the exact ones.
In Fig.~\ref{fig:metric}, we have computed the values of $g_{00}$ and $g_{11}$ on 100 new points for each sub-domain, using the trained deep neural networks. The results are concatenated for visualization.

\begin{figure}
    \centering \includegraphics[width=1\textwidth]{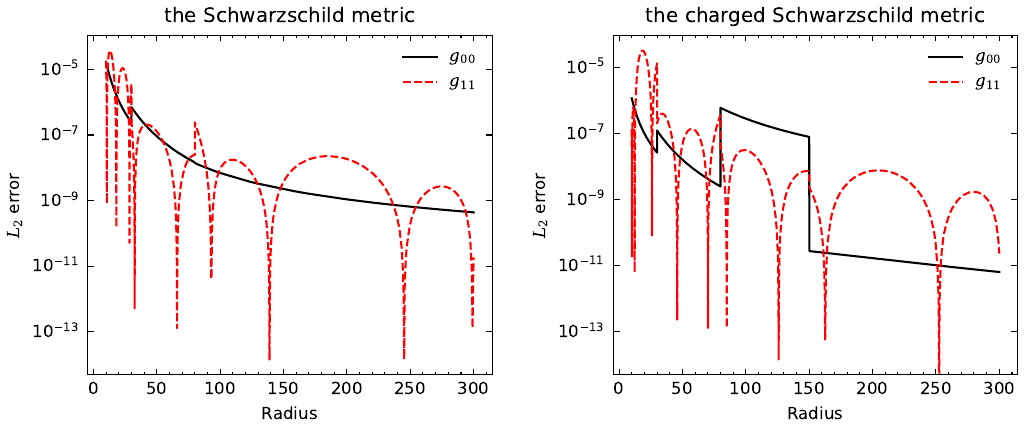}
    \caption{Left: $L_2$ error between the predicted and exact 
    solution of the Schwarzschild metric. Right: $L_2$ error between the predicted and exact solution of the charged Schwarzschild metric.}
    \label{fig:L2error}
\end{figure}

To quantify the difference between the network prediction and the analytical solution, we computed the $L_2$ error averaged over 1024 points, defined as $L_2 = {1 \over 1024 }\sum_{i=1}^{1024} (g_{\mu\nu}^{\rm net} - g_{\mu\nu}^{\rm exact})_i^2$. 
Fig.~\ref{fig:L2error} shows the $L_2$ error for the Schwarzschild metric (left) and the charged Schwarzschild metric (right).
Both errors decrease as the $r$ increases, in general.
The $g_{00}$ in both cases are much smoother than the learned $g_{11}$, which agrees with our experience that the $g_{11}$ converges much slower than the $g_{00}$.
In the Schwarzschild metric, the learned $g_{00}$ shows one visible discontinuity at $r=30$ and one negligible discontinuity at $r=80$.
In the charged Schwarzschild metric, the learned $g_{00}$ shows 3 discontinuities between each pair of contacting sub-domains.
In both cases, the $g_{11}$ shows many discontinuities besides the interface of sub-domains.
The maximum error of the metric component between prediction and the exact solution is approximately $10^{-3}$, it decreases to about $10^{-5}$ at $r\rightarrow 300$.



\section{Summary}


We develop a new method to solve the Einstein field equations numerically using deep learning and auto differentiation. The method is used to extract the Schwarzschild metric field and the charged Schwarzschild metric field given the matter distribution.
The maximum relative error between the numerical and the analytical solution is as small as $10^{-3}$.
In this method, we use several physically constrained neural networks to represent metric fields at different space-time regions. The network is constructed to obey boundary conditions at $r\rightarrow \infty$ naturally. 
We use auto differentiation to compute the derivatives of the metric fields, the Christoffel symbols and the Ricci tensor with regard to $x^{\mu}$.
The violations of the neural network solution to the Einstein field equations are used as training objectives during optimization. 

Compared to the traditional numerical relativity, the present method is mesh-free.
We do not need to approximate various derivatives in the Einstein field equations using finite difference method on regular space-time grids, 
because the auto differentiation brings analytical precision.
The numerical error of the solution is controllable by adding more testing points in the given regions or more neurons in the hidden layers of the network.
The problem of solving PDEs is translated into a problem of optimization, 
which is much more stable numerically as long as the network has enough representation power and the testing points are rich enough. 
We noticed that \cite{Yan:2020wcd} use a deep neural network to learn the black hole metrics from frequency-dependent shear viscosity and \cite{Li:2022zjc, Xu:2023eof} use other methods baesd on deep learning in AdS/CFT correspondence.
However, they have not used the PINN method and the auto differentiation to compute the derivatives in the Ricci tensor.
To learn the black hole metric, they have to prepare supervised data assuming the metric is known in the forward process.
Our method belongs to unsupervised learning that do not need labelled data. 
And \cite{Luna:2022rql} utilized the PINN method to learn Kerr metric from Teukolsky equation , but not from the Einstein Field Equation directly.

However, like other deep learning research, our method suffers from a common problem of hyperparameter tuning - how to find the optimal hyperparameters to minimize the violation of the metric fields to the Einstein field equations. To get the current results, we observe that LogSigmoid activation, DPINN and the physically constrained network structure are important for fast training. These experience should be important for the future studies of numerical relativity using deep learning.
In addition, we have not yet demonstrated the effectiveness of this method in other geometries, especially the space-time coupled geometry. 
In the future, we hope to generalize this method to more situations, such as the time evolution problems under non-static, non-symmetric matter distribution, which can give us a better understanding of various physical phenomena in the universe that are affected by general relativity. Since computing the induced metric $\gamma_{ij}$ and extrinsic curvature $K_{ij}$ from the energy-momentum tensor and the metric field is not a straightforward task, if we can get suitable munerical format, we expect to be able to treat the Einstein field equations as ordinary partial differential equations through puting physical constrains (such as coordinate gauge) into neural network. However, by providing the distribution of matter and, if necessary, the metric information at initial time, this method cantnot avoid the explicit computation of the induced metric and the extrinsic curvature to solve for the evolution of metric field directly because the form of the Einstein field equation (Eq.~\ref{Einstein_field_equation}) may cause error accumulation.

In nuclear physics, the mass radius relation of neutron stars is widely used to extract the nuclear Equation of State (EoS) at high density. PINN was used to represent the unknown nuclear EoS and to solve the TOV equation, which helps to extract the nuclear EoS using observed mass radius data \cite{Zhou:2023pti,Boehnlein:2021eym,Fujimoto:2019hxv,1711.06748}. In their studies, the metric fields do not depend on the matter distribution inside the neutron stars. In principle, we can combine our methods with the TOV solver, to determine the nuclear EoS more consistently.
When using PINNs to solve PDEs, it is not necessary to know the specific properties of the problem in advance, and it can handle different types of physical problems, even with electromagnetic interaction.
Our method may pave a new way in the related issues which need to solve the metric field.

\begin{acknowledgments}
This work was supported by the National Science Foundation of China under Grant Nos. 12075098. 
\end{acknowledgments}

\bibliographystyle{unsrt}
\bibliography{ref.bib}

\begin{thebibliography}{100}

\bibitem{HAHN1964304}
Susan~G Hahn and Richard~W Lindquist.
\newblock The two-body problem in geometrodynamics.
\newblock {\em Annals of Physics}, 29(2):304--331, 1964.

\bibitem{1975PhDT1}
K.~R. {Eppley}.
\newblock {\em {The numerical evolution of the collision of two black holes}}.
\newblock PhD thesis, Princeton University, New Jersey, January 1975.

\bibitem{1975PhDT2}
L.~L. {Smarr}.
\newblock {\em {The structure of general relativity with a numerical
  illustration: The collision of two black holes}}.
\newblock PhD thesis, University of Texas, Austin, January 1975.

\bibitem{tb37076x}
Larry Smarr.
\newblock Space-times generated by computers: Black holes with gravitational
  radiation*.
\newblock {\em Annals of the New York Academy of Sciences}, 302(1):569--604,
  1977.

\bibitem{PhysRevD.14.2443}
Larry Smarr, Andrej \ifmmode \check{C}\else
  \v{C}\fi{}ade\ifmmode~\check{z}\else \v{z}\fi{}, Bryce DeWitt, and Kenneth
  Eppley.
\newblock Collision of two black holes: Theoretical framework.
\newblock {\em Phys. Rev. D}, 14:2443--2452, Nov 1976.

\bibitem{PhysRevLett.41.1085}
Tsvi Piran.
\newblock Cylindrical general relativistic collapse.
\newblock {\em Phys. Rev. Lett.}, 41:1085--1088, Oct 1978.

\bibitem{PhysRevLett.55.891}
Richard~F. Stark and Tsvi Piran.
\newblock Gravitational-wave emission from rotating gravitational collapse.
\newblock {\em Phys. Rev. Lett.}, 55:891--894, Aug 1985.

\bibitem{19901}
Peter Anninos, Joan Mass\'o, Edward Seidel, Wai-Mo Suen, and John Towns.
\newblock Three-dimensional numerical relativity: The evolution of black holes.
\newblock {\em Phys. Rev. D}, 52:2059--2082, Aug 1995.

\bibitem{19902}
R.~G\'omez, L.~Lehner, and Marsa .ect.
\newblock Stable characteristic evolution of generic three-dimensional
  single-black-hole spacetimes.
\newblock {\em Phys. Rev. Lett.}, 80:3915--3918, May 1998.

\bibitem{19903}
Bernd Bruegmann.
\newblock Binary black hole mergers in 3d numerical relativity.
\newblock 1997.

\bibitem{19904}
Lawrence~E. Kidder, Mark~A. Scheel, Saul~A. Teukolsky, Eric~D. Carlson, and
  Gregory~B. Cook.
\newblock Black hole evolution by spectral methods.
\newblock {\em Phys. Rev. D}, 62:084032, Sep 2000.

\bibitem{19905}
Gregory~B. Cook.
\newblock Corotating and irrotational binary black holes in quasicircular
  orbits.
\newblock {\em Phys. Rev. D}, 65:084003, Mar 2002.

\bibitem{19906}
Gregory~B. Cook.
\newblock Three-dimensional initial data for the collision of two black holes.
  ii. quasicircular orbits for equal-mass black holes.
\newblock {\em Phys. Rev. D}, 50:5025--5032, Oct 1994.

\bibitem{19907}
Gregory~B. Cook and Harald~P. Pfeiffer.
\newblock Excision boundary conditions for black-hole initial data.
\newblock {\em Phys. Rev. D}, 70:104016, Nov 2004.

\bibitem{19908}
Thomas~W. Baumgarte.
\newblock Innermost stable circular orbit of binary black holes.
\newblock {\em Phys. Rev. D}, 62:024018, Jun 2000.

\bibitem{19909}
Steven Brandt and Bernd Br\"ugmann.
\newblock A simple construction of initial data for multiple black holes.
\newblock {\em Phys. Rev. Lett.}, 78:3606--3609, May 1997.

\bibitem{199010}
Edward Seidel and Wai-Mo Suen.
\newblock Towards a singularity-proof scheme in numerical relativity.
\newblock {\em Phys. Rev. Lett.}, 69:1845--1848, Sep 1992.

\bibitem{199011}
Peter Anninos, Greg Daues, Joan Mass\'o, Edward Seidel, and Wai-Mo Suen.
\newblock Horizon boundary condition for black hole spacetimes.
\newblock {\em Phys. Rev. D}, 51:5562--5578, May 1995.

\bibitem{199012}
Miguel Alcubierre and Bernd Br\"ugmann.
\newblock Simple excision of a black hole in $3+1$ numerical relativity.
\newblock {\em Phys. Rev. D}, 63:104006, Apr 2001.

\bibitem{199013}
Helmut Friedrich and Alan Rendall.
\newblock The cauchy problem for the einstein equations.
\newblock In Bernd~G. Schmidt, editor, {\em Einstein's Field Equations and
  Their Physical Implications}, pages 127--223, Berlin, Heidelberg, 2000.
  Springer Berlin Heidelberg.

\bibitem{199014}
C.~Bona and J.~Mass\'o.
\newblock Hyperbolic evolution system for numerical relativity.
\newblock {\em Phys. Rev. Lett.}, 68:1097--1099, Feb 1992.

\bibitem{199015}
D.~Shoemaker, K.~Smith, U.~Sperhake, P.~Laguna, E.~Schnetter, and D.~Fiske.
\newblock Moving black holes via singularity excision.
\newblock 2003.

\bibitem{199016}
Carles Bona, Joan Mass\'o, Edward Seidel, and Joan Stela.
\newblock New formalism for numerical relativity.
\newblock {\em Phys. Rev. Lett.}, 75:600--603, Jul 1995.

\bibitem{199017}
Andrew Abrahams, Arlen Anderson, Yvonne Choquet-Bruhat, and Jr~James W.~York.
\newblock Hyperbolic formulation of general relativity, 1997.

\bibitem{199018}
Takashi Nakamura, Kenichi Oohara, and Yasufumi Kojima.
\newblock {General Relativistic Collapse to Black Holes and Gravitational Waves
  from Black Holes}.
\newblock {\em Progress of Theoretical Physics Supplement}, 90:1--218, 01 1987.

\bibitem{199019}
Masaru Shibata and Takashi Nakamura.
\newblock Evolution of three-dimensional gravitational waves: Harmonic slicing
  case.
\newblock {\em Phys. Rev. D}, 52:5428--5444, Nov 1995.

\bibitem{199020}
Thomas~W. Baumgarte and Stuart~L. Shapiro.
\newblock Numerical integration of einstein's field equations.
\newblock {\em Phys. Rev. D}, 59:024007, Dec 1998.

\bibitem{199021}
Karen Camarda and Edward Seidel.
\newblock Three-dimensional simulations of distorted black holes: Comparison
  with axisymmetric results.
\newblock {\em Phys. Rev. D}, 59:064019, Feb 1999.

\bibitem{199022}
Ulrich Sperhake, Bernard Kelly, Pablo Laguna, Kenneth~L. Smith, and Erik
  Schnetter.
\newblock Black hole head-on collisions and gravitational waves with fixed
  mesh-refinement and dynamic singularity excision.
\newblock {\em Phys. Rev. D}, 71:124042, Jun 2005.

\bibitem{199023}
Bernd Bruegmann.
\newblock {Binary black hole mergers in 3-d numerical relativity}.
\newblock {\em Int. J. Mod. Phys. D}, 8:85, 1999.

\bibitem{199024}
Steve Brandt and Correll \etal.
\newblock Grazing collisions of black holes via the excision of singularities.
\newblock {\em Phys. Rev. Lett.}, 85:5496--5499, Dec 2000.

\bibitem{199025}
Miguel Alcubierre, Werner Benger, Bernd Br\"ugmann, Gerd Lanfermann, Lars
  Nerger, Edward Seidel, and Ryoji Takahashi.
\newblock 3d grazing collision of two black holes.
\newblock {\em Phys. Rev. Lett.}, 87:271103, Dec 2001.

\bibitem{199026}
Miguel Alcubierre, Bernd Br\"ugmann, Peter Diener, Michael Koppitz, Denis
  Pollney, Edward Seidel, and Ryoji Takahashi.
\newblock Gauge conditions for long-term numerical black hole evolutions
  without excision.
\newblock {\em Phys. Rev. D}, 67:084023, Apr 2003.

\bibitem{199027}
Peter MacNeice, Kevin~M. Olson, Clark Mobarry, Rosalinda {de Fainchtein}, and
  Charles Packer.
\newblock Paramesh: A parallel adaptive mesh refinement community toolkit.
\newblock {\em Computer Physics Communications}, 126(3):330--354, 2000.

\bibitem{199028}
Harald~P. Pfeiffer, Lawrence~E. Kidder, Mark~A. Scheel, and Saul~A. Teukolsky.
\newblock A multidomain spectral method for solving elliptic equations.
\newblock {\em Computer Physics Communications}, 152(3):253--273, 2003.

\bibitem{199029}
J.~Baker, B.~Bruegmann, M.~Campanelli, and C.~O. Lousto.
\newblock Gravitational waves from black hole collisions via an eclectic
  approach.
\newblock 2000.

\bibitem{199030}
J.~Baker, B.~Br\"ugmann, M.~Campanelli, C.~O. Lousto, and R.~Takahashi.
\newblock Plunge waveforms from inspiralling binary black holes.
\newblock {\em Phys. Rev. Lett.}, 87:121103, Aug 2001.

\bibitem{199031}
John Baker, Manuela Campanelli, and Carlos~O. Lousto.
\newblock The lazarus project: A pragmatic approach to binary black hole
  evolutions.
\newblock {\em Phys. Rev. D}, 65:044001, Jan 2002.

\bibitem{199032}
J.~Baker, M.~Campanelli, C.~O. Lousto, and R.~Takahashi.
\newblock Modeling gravitational radiation from coalescing binary black holes.
\newblock {\em Phys. Rev. D}, 65:124012, Jun 2002.

\bibitem{199033}
S.~R. Brandt, K.~Camarda, and E.~Seidel.
\newblock Three dimensional distorted black holes: Initial data and evolution,
  1997.

\bibitem{199034}
G.~B. Cook, M.~F. Huq, and S.~A~\etal Klasky.
\newblock Boosted three-dimensional black-hole evolutions with singularity
  excision.
\newblock {\em Phys. Rev. Lett.}, 80:2512--2516, Mar 1998.

\bibitem{199035}
Bernd Br\"ugmann, Wolfgang Tichy, and Nina Jansen.
\newblock Numerical simulation of orbiting black holes.
\newblock {\em Phys. Rev. Lett.}, 92:211101, May 2004.

\bibitem{PhysRevLett.95.121101}
Frans Pretorius.
\newblock Evolution of binary black-hole spacetimes.
\newblock {\em Phys. Rev. Lett.}, 95:121101, Sep 2005.

\bibitem{PhysRevLett.96.111102}
John~G. Baker, Joan Centrella, Dae-Il Choi, Michael Koppitz, and James van
  Meter.
\newblock Gravitational-wave extraction from an inspiraling configuration of
  merging black holes.
\newblock {\em Phys. Rev. Lett.}, 96:111102, Mar 2006.

\bibitem{PhysRevLett.96.111101}
M.~Campanelli, C.~O. Lousto, P.~Marronetti, and Y.~Zlochower.
\newblock Accurate evolutions of orbiting black-hole binaries without excision.
\newblock {\em Phys. Rev. Lett.}, 96:111101, Mar 2006.

\bibitem{Centrella:2010mx}
Joan Centrella, John~G. Baker, Bernard~J. Kelly, and James~R. van Meter.
\newblock {Black-hole binaries, gravitational waves, and numerical relativity}.
\newblock {\em Rev. Mod. Phys.}, 82:3069, 2010.

\bibitem{science361}
Bernd Brügmann.
\newblock Fundamentals of numerical relativity for gravitational wave sources.
\newblock {\em Science}, 361(6400):366--371, 2018.

\bibitem{Kessler:2016uwi}
Richard Kessler and Dan Scolnic.
\newblock {Correcting Type Ia Supernova Distances for Selection Biases and
  Contamination in Photometrically Identified Samples}.
\newblock {\em Astrophys. J.}, 836(1):56, 2017.

\bibitem{DES:2015bqp}
C.~Bonnett et~al.
\newblock {Redshift distributions of galaxies in the Dark Energy Survey Science
  Verification shear catalogue and implications for weak lensing}.
\newblock {\em Phys. Rev. D}, 94(4):042005, 2016.

\bibitem{Sanders:2014uva}
N.~E. Sanders et~al.
\newblock {Towards Characterization of the Type IIP Supernova Progenitor
  Population: a Statistical Sample of Light Curves from Pan-STARRS1}.
\newblock {\em Astrophys. J.}, 799(2):208, 2015.

\bibitem{Wong:2020ise}
Kaze W.~K. Wong, Katelyn Breivik, Kyle Kremer, and Thomas Callister.
\newblock {Joint constraints on the field-cluster mixing fraction, common
  envelope efficiency, and globular cluster radii from a population of binary
  hole mergers via deep learning}.
\newblock {\em Phys. Rev. D}, 103(8):083021, 2021.

\bibitem{Lanusse:2017vha}
Francois Lanusse, Quanbin Ma, Nan Li, Thomas~E. Collett, Chun-Liang Li, Siamak
  Ravanbakhsh, Rachel Mandelbaum, and Barnabas Poczos.
\newblock {CMU DeepLens: deep learning for automatic image-based
  galaxy\textendash{}galaxy strong lens finding}.
\newblock {\em Mon. Not. Roy. Astron. Soc.}, 473(3):3895--3906, 2018.

\bibitem{PerreaultLevasseur:2017ltk}
Laurence Perreault~Levasseur, Yashar~D. Hezaveh, and Risa~H. Wechsler.
\newblock {Uncertainties in Parameters Estimated with Neural Networks:
  Application to Strong Gravitational Lensing}.
\newblock {\em Astrophys. J. Lett.}, 850(1):L7, 2017.

\bibitem{Fluri:2018hoy}
Janis Fluri, Tomasz Kacprzak, Alexandre Refregier, Adam Amara, Aurelien Lucchi,
  and Thomas Hofmann.
\newblock {Cosmological constraints from noisy convergence maps through deep
  learning}.
\newblock {\em Phys. Rev. D}, 98(12):123518, 2018.

\bibitem{George:2017pmj}
Daniel George and E.~A. Huerta.
\newblock {Deep Learning for Real-time Gravitational Wave Detection and
  Parameter Estimation: Results with Advanced LIGO Data}.
\newblock {\em Phys. Lett. B}, 778:64--70, 2018.

\bibitem{Zevin:2016qwy}
Michael Zevin et~al.
\newblock {Gravity Spy: Integrating Advanced LIGO Detector Characterization,
  Machine Learning, and Citizen Science}.
\newblock {\em Class. Quant. Grav.}, 34(6):064003, 2017.

\bibitem{George:2016hay}
Daniel George and E.~A. Huerta.
\newblock {Deep Neural Networks to Enable Real-time Multimessenger
  Astrophysics}.
\newblock {\em Phys. Rev. D}, 97(4):044039, 2018.

\bibitem{Wong:2020yig}
Kaze W.~K. Wong, Gabriele Franciolini, Valerio De~Luca, Vishal Baibhav,
  Emanuele Berti, Paolo Pani, and Antonio Riotto.
\newblock {Constraining the primordial black hole scenario with Bayesian
  inference and machine learning: the GWTC-2 gravitational wave catalog}.
\newblock {\em Phys. Rev. D}, 103(2):023026, 2021.

\bibitem{Green:2020hst}
Stephen~R. Green, Christine Simpson, and Jonathan Gair.
\newblock {Gravitational-wave parameter estimation with autoregressive neural
  network flows}.
\newblock {\em Phys. Rev. D}, 102(10):104057, 2020.

\bibitem{George:2018awu}
Daniel George, Hongyu Shen, and E.~A. Huerta.
\newblock {Classification and unsupervised clustering of LIGO data with Deep
  Transfer Learning}.
\newblock {\em Phys. Rev. D}, 97(10):101501, 2018.

\bibitem{Vajente:2019ycy}
Gabriele Vajente, Yiwen Huang, Maximiliano Isi, Jenne~C. Driggers, Jeffrey~S.
  Kissel, Marek~J. Szczepanczyk, and Salvatore Vitale.
\newblock {Machine-learning nonstationary noise out of gravitational-wave
  detectors}.
\newblock {\em Phys. Rev. D}, 101(4):042003, 2020.

\bibitem{Chua:2019wwt}
Alvin J.~K. Chua and Michele Vallisneri.
\newblock {Learning Bayesian posteriors with neural networks for
  gravitational-wave inference}.
\newblock {\em Phys. Rev. Lett.}, 124(4):041102, 2020.

\bibitem{Razzano:2018fxb}
Massimiliano Razzano and Elena Cuoco.
\newblock {Image-based deep learning for classification of noise transients in
  gravitational wave detectors}.
\newblock {\em Class. Quant. Grav.}, 35(9):095016, 2018.

\bibitem{Wei:2019zlc}
Wei Wei and E.~A. Huerta.
\newblock {Gravitational Wave Denoising of Binary Black Hole Mergers with Deep
  Learning}.
\newblock {\em Phys. Lett. B}, 800:135081, 2020.

\bibitem{Dax:2021tsq}
Maximilian Dax, Stephen~R. Green, Jonathan Gair, Jakob~H. Macke, Alessandra
  Buonanno, and Bernhard Sch\"olkopf.
\newblock {Real-Time Gravitational Wave Science with Neural Posterior
  Estimation}.
\newblock {\em Phys. Rev. Lett.}, 127(24):241103, 2021.

\bibitem{Shen:2019vep}
Hongyu Shen, E.~A. Huerta, Eamonn O'Shea, Prayush Kumar, and Zhizhen Zhao.
\newblock {Statistically-informed deep learning for gravitational wave
  parameter estimation}.
\newblock {\em Mach. Learn. Sci. Tech.}, 3(1):015007, 2022.

\bibitem{1711.06464}
Jens Berg and Kaj Nyström.
\newblock A unified deep artificial neural network approach to partial
  differential equations in complex geometries.
\newblock 2017.

\bibitem{2}
M.~Raissi, P.~Perdikaris, and G.E. Karniadakis.
\newblock Physics-informed neural networks: A deep learning framework for
  solving forward and inverse problems involving nonlinear partial differential
  equations.
\newblock {\em Journal of Computational Physics}, 378:686--707, 2019.

\bibitem{PINN}
George Karniadakis, Yannis Kevrekidis, Lu~Lu, Paris Perdikaris, Sifan Wang, and
  Liu Yang.
\newblock Physics-informed machine learning.
\newblock pages 1--19, 05 2021.

\bibitem{Karniadakis2021-nq}
George~Em Karniadakis, Ioannis~G Kevrekidis, Lu~Lu, Paris Perdikaris, Sifan
  Wang, and Liu Yang.
\newblock Physics-informed machine learning.
\newblock {\em Nature Reviews Physics}, 3(6):422--440, June 2021.

\bibitem{2107.09443}
Kirill Zubov, Zoe McCarthy, Yingbo Ma, Francesco Calisto, Valerio Pagliarino,
  Simone Azeglio, Luca Bottero, Emmanuel Luján, Valentin Sulzer, Ashutosh
  Bharambe, Nand Vinchhi, Kaushik Balakrishnan, Devesh Upadhyay, and Chris
  Rackauckas.
\newblock Neuralpde: Automating physics-informed neural networks (pinns) with
  error approximations, 2021.

\bibitem{JPSJ.87.074002}
Hiroki Saito.
\newblock Method to solve quantum few-body problems with artificial neural
  networks.
\newblock {\em Journal of the Physical Society of Japan}, 87(7):074002, 2018.

\bibitem{KEEBLE2020135743}
J.W.T. Keeble and A.~Rios.
\newblock Machine learning the deuteron.
\newblock {\em Physics Letters B}, 809:135743, 2020.

\bibitem{1909.08423}
Jan Hermann, Zeno Schätzle, and Frank Noé.
\newblock Deep neural network solution of the electronic schrödinger equation.
\newblock 2019.

\bibitem{sci}
Giuseppe Carleo and Matthias Troyer.
\newblock Solving the quantum many-body problem with artificial neural
  networks.
\newblock {\em Science}, 355(6325):602--606, 2017.

\bibitem{Pu2023jae}
Kai-Fang Pu, Hanlin Li, Hong-Liang Lu, and Long-Gang Pang.
\newblock {Solving Schrodinger equations using a physically constrained neural
  network*}.
\newblock {\em Chin. Phys. C}, 47(5):054104, 2023.

\bibitem{Adams:2020aax}
Corey Adams, Giuseppe Carleo, Alessandro Lovato, and Noemi Rocco.
\newblock {Variational Monte Carlo Calculations of A\ensuremath{\leq}4 Nuclei
  with an Artificial Neural-Network Correlator Ansatz}.
\newblock {\em Phys. Rev. Lett.}, 127(2):022502, 2021.

\bibitem{Shi:2021qri}
Shuzhe Shi, Kai Zhou, Jiaxing Zhao, Swagato Mukherjee, and Pengfei Zhuang.
\newblock {Heavy quark potential in the quark-gluon plasma: Deep neural network
  meets lattice quantum chromodynamics}.
\newblock {\em Phys. Rev. D}, 105(1):014017, 2022.

\bibitem{Yang:2022rlw}
Yilong Yang and Pengwei Zhao.
\newblock {Deep-neural-network approach to solving the ab initio nuclear
  structure problem}.
\newblock {\em Phys. Rev. C}, 107(3):034320, 2023.

\bibitem{LEUNG2022111539}
Wing~Tat Leung, Guang Lin, and Zecheng Zhang.
\newblock Nh-pinn: Neural homogenization-based physics-informed neural network
  for multiscale problems.
\newblock {\em Journal of Computational Physics}, 470:111539, 2022.

\bibitem{2109.09444}
Zheyuan Hu, Ameya~D. Jagtap, George~Em Karniadakis, and Kenji Kawaguchi.
\newblock When do extended physics-informed neural networks (xpinns) improve
  generalization?
\newblock 2021.

\bibitem{2002.08235}
Teeratorn Kadeethum, Thomas~M Jorgensen, and Hamidreza~M Nick.
\newblock Physics-informed neural networks for solving nonlinear diffusivity
  and biot's equations.
\newblock 2020.

\bibitem{2304.01670}
Duo Xu, Jonathan~C. Tan, Chia-Jung Hsu, and Ye~Zhu.
\newblock Denoising diffusion probabilistic models to predict the density of
  molecular clouds, 2023.

\bibitem{2010.08895}
Zongyi Li, Nikola Kovachki, Kamyar Azizzadenesheli, Burigede Liu, Kaushik
  Bhattacharya, Andrew Stuart, and Animashree Anandkumar.
\newblock Fourier neural operator for parametric partial differential
  equations.
\newblock 10 2020.

\bibitem{Rosofsky:2023dtc}
Shawn~G. Rosofsky and E.~A. Huerta.
\newblock {Magnetohydrodynamics with Physics Informed Neural Operators}.
\newblock 2 2023.

\bibitem{2305.07940}
Xiaofei Guan, Boya Hu, Shipeng Mao, and Xintong Wang.
\newblock Mhdnet: Multi-modes multiscale physics informed neural networks for
  solving magnetohydrodynamics problems, 2023.

\bibitem{Rosofsky:2022lgb}
Shawn~G. Rosofsky, Hani~Al Majed, and E.~A. Huerta.
\newblock {Applications of physics informed neural operators}.
\newblock {\em Mach. Learn. Sci. Tech.}, 4(2):025022, 2023.

\bibitem{1912.11073}
Shawn~G. Rosofsky and E.~A. Huerta.
\newblock Artificial neural network subgrid models of 2-d compressible
  magnetohydrodynamic turbulence.
\newblock 2019.

\bibitem{2211.14266}
Duo Xu, Chi-Yan Law, and Jonathan~C. Tan.
\newblock Application of convolutional neural networks to predict magnetic
  fields directions in turbulent clouds.
\newblock 2022.

\bibitem{2102.01447}
Panagiotis Tigas, Téo Bloch, Vishal Upendran, Banafsheh Ferdoushi, Mark C.~M.
  Cheung, Siddha Ganju, Ryan~M. McGranaghan, Yarin Gal, and Asti Bhatt.
\newblock Global earth magnetic field modeling and forecasting with spherical
  harmonics decomposition, 2021.

\bibitem{Karpov:2022tro}
Platon~I. Karpov, Chengkun Huang, Iskandar Sitdikov, Chris~L. Fryer, Stan
  Woosley, and Ghanshyam Pilania.
\newblock {Physics-informed Machine Learning for Modeling Turbulence in
  Supernovae}.
\newblock {\em Astrophys. J.}, 940(1):26, 2022.

\bibitem{2211.08579}
Souvik Bose, Bart~De Pontieu, Viggo Hansteen, Alberto~Sainz Dalda, Sabrina
  Savage, and Amy Winebarger.
\newblock Chromospheric and coronal heating in active region plage by
  dissipation of currents from braiding, 2022.

\bibitem{1502.05767}
Atilim~Gunes Baydin, Barak~A. Pearlmutter, Alexey~Andreyevich Radul, and
  Jeffrey~Mark Siskind.
\newblock Automatic differentiation in machine learning: a survey.
\newblock 2015.

\bibitem{1412.6980}
Diederik~P. Kingma and Jimmy Ba.
\newblock Adam: A method for stochastic optimization, 2014.

\bibitem{1912.01703}
Adam Paszke, Sam Gross, Francisco Massa, Adam Lerer, James Bradbury, Gregory
  Chanan, Trevor Killeen, Zeming Lin, Natalia Gimelshein, Luca Antiga, Alban
  Desmaison, Andreas Köpf, Edward Yang, Zach DeVito, Martin Raison, Alykhan
  Tejani, Sasank Chilamkurthy, Benoit Steiner, Lu~Fang, Junjie Bai, and Soumith
  Chintala.
\newblock Pytorch: An imperative style, high-performance deep learning library,
  2019.

\bibitem{1907.08967}
Vikas Dwivedi, Nishant Parashar, and Balaji Srinivasan.
\newblock Distributed physics informed neural network for data-efficient
  solution to partial differential equations, 2019.

\bibitem{article111}
Ameya Jagtap and George Karniadakis.
\newblock Extended physics-informed neural networks (xpinns): A generalized
  space-time domain decomposition based deep learning framework for nonlinear
  partial differential equations.
\newblock {\em Communications in Computational Physics}, 28:2002--2041, 11
  2020.

\bibitem{JAGTAP2020113028}
Ameya~D. Jagtap, Ehsan Kharazmi, and George~Em Karniadakis.
\newblock Conservative physics-informed neural networks on discrete domains for
  conservation laws: Applications to forward and inverse problems.
\newblock {\em Computer Methods in Applied Mechanics and Engineering},
  365:113028, 2020.

\bibitem{Yan:2020wcd}
Yu-Kun Yan, Shao-Feng Wu, Xian-Hui Ge, and Yu~Tian.
\newblock {Deep learning black hole metrics from shear viscosity}.
\newblock {\em Phys. Rev. D}, 102(10):101902, 4 2020.

\bibitem{Li:2022zjc}
Kai Li, Yi~Ling, Peng Liu, and Meng-He Wu.
\newblock {Learning the black hole metric from holographic conductivity}.
\newblock {\em Phys. Rev. D}, 107(6):066021, 2023.

\bibitem{Xu:2023eof}
Wen-Bin Xu and Shao-Feng Wu.
\newblock {Reconstructing black hole exteriors and interiors using entanglement
  and complexity}.
\newblock 5 2023.

\bibitem{Luna:2022rql}
Raimon Luna, Juan Calder\'on~Bustillo, Juan Jos\'e~Seoane Mart\'\i{}nez,
  Alejandro Torres-Forn\'e, and Jos\'e~A. Font.
\newblock {Solving the Teukolsky equation with physics-informed neural
  networks}.
\newblock {\em Phys. Rev. D}, 107(6):064025, 2023.

\bibitem{Zhou:2023pti}
Kai Zhou, Lingxiao Wang, Long-Gang Pang, and Shuzhe Shi.
\newblock {Exploring QCD matter in extreme conditions with Machine Learning}.
\newblock 3 2023.

\bibitem{Boehnlein:2021eym}
Amber Boehnlein et~al.
\newblock {Colloquium: Machine learning in nuclear physics}.
\newblock {\em Rev. Mod. Phys.}, 94(3):031003, 2022.

\bibitem{Fujimoto:2019hxv}
Yuki Fujimoto, Kenji Fukushima, and Koichi Murase.
\newblock {Mapping neutron star data to the equation of state using the deep
  neural network}.
\newblock {\em Phys. Rev. D}, 101(5):054016, 2020.

\bibitem{1711.06748}
Yuki Fujimoto, Kenji Fukushima, and Koichi Murase.
\newblock Methodology study of machine learning for the neutron star equation
  of state.
\newblock {\em Phys. Rev. D}, 2017.

\end{thebibliography}

\end{document}